\documentclass[12pt]{template}

\usepackage{graphicx}
\usepackage{subcaption}
\usepackage{fancyhdr}
\AtBeginDocument{%
  \pagestyle{fancy}  
  \fancyhf{}         
  \fancyfoot[C]{\thepage}  
  \setlength{\footskip}{20mm}  
  \maketitle
  \let\maketitle\relax
}

\title{Deep Reinforcement Learning for Active Flow Control around a Three-Dimensional Flow-Separated Wing at $Re = 1,000$}

\author{R. Montalà$^{1,*}$, B. Font$^{2}$, P. Suárez$^{3}$, J. Rabault$^{4}$, O. Lehmkuhl$^{5}$, R. Vinuesa$^{3}$ and I. Rodriguez$^{1}$}

\address{$^{1}$ TUAREG, Universitat Politècnica de Catalunya (UPC), Spain
\and
$^{2}$ Mechanical Engineering, Delft University of Technology (TU Delft), Netherlands
\and
$^{3}$ FLOW, Engineering Mechanics, KTH Royal Institute of Technology, Sweden
\and
$^{4}$ Independent Researcher, Oslo, Norway
\and
$^{5}$ LS/CFD - CASE, Barcelona Supercomputing Center (BSC), Spain
\and
$^{*}$ ricard.montala@upc.edu}

\begin{document}

\vskip0.5cm

\noindent {\bf Abstract}: This study explores the use of deep reinforcement learning (DRL) for active flow control (AFC) to reduce flow separation on wings at high angles of attack. Concretely, here the DRL agent controls the flow over the three-dimensional NACA0012 wing section at the Reynolds number $Re = 1,000$ and angle of attack AoA$ = 20^\circ$, autonomously identifying optimal control actions through real-time flow data and a reward function focused on improving aerodynamic performance. The framework integrates the GPU-accelerated computational fluid dynamics (CFD) solver SOD2D with the TF-Agents DRL library via a Redis in-memory database, enabling rapid training. This work builds on previous DRL flow-control studies, demonstrating DRL’s potential to address complex aerodynamic challenges and push the boundaries of traditional AFC methods.

\vskip0.5cm

\section{Introduction}
Active flow control (AFC) has emerged as a potential solution to enhance aerodynamic performance and mitigate climate change in the transportation sector. By applying this technique to bluff bodies such as road vehicles \cite{Cerutti2020}, trucks \cite{Minelli2016}, or airfoils at high angles of attack \cite{Rodriguez2020}, drag can be minimized, hence improving fuel efficiency and reducing CO$_2$ emissions.

Over the years, several studies have explored various actuation laws to minimize drag, including constant blowing, constant suction, or harmonic actuations such as synthetic jets. However, the rapid advancement of machine learning (ML), particularly deep reinforcement learning (DRL), has introduced a completely new approach to flow control \cite{Garnier2024}. By leveraging this powerful data-driven tool with fluid dynamics, a ML model can be developed to automatically optimize jet actuation based on real-time flow conditions, i.e., closed-loop control. Beyond the benefits of closed-loop control, DRL models can identify the complex interactions between the jets and the flow during the training, enabling richer interactions with the flow as the training advances, achieving greater performance gains compared to traditional methods, where normally a single jet actuation frequency is targeted.

Recent studies have demonstrated the potential of DRL in AFC problems, beginning with the pioneering work of Rabault et al. \cite{Rabault2019a}, who achieved drag reduction on a two-dimensional cylinder at $Re = 100$. Subsequent research extended DRL applications to higher Reynolds numbers, with improvements in training efficiency through multi-environment approaches \cite{Rabault2019b}. Notably, Suárez et al. extended the methodology to a three-dimensional flow \cite{Suarez2025a} and even to a Reynolds number of $Re=3,900$ \cite{Suarez2025b}, achieving significant drag reductions. Other studies have applied DRL to reducing skin friction in wall-bounded flows at low Reynolds numbers, i.e., $Re_\tau = 150-200$ \cite{Guastoni2023, Sonoda2023, Cavallazzi2024}, or controlling the three-dimensional Rayleigh–Bénard convection \cite{Vasanth2024}. Some DRL applications for AFC on wings can also be found in the literature \cite{Wang2022,Garcia2025}. However, these works are limitted to two-dimensional airfoils. Therefore, despite these advancements, the combination of DRL with AFC for flow control is still in its early stages of development, and this methodology remains constrained to low Reynolds numbers and/or canonical cases, limiting its industrial relevance. 

A key challenge in DRL-based AFC is the “two-language” problem: Normally, ML libraries use high-level languages like Python, while physics solvers rely on low-level languages like C++ or Fortran. Efficiently integrating these two components is critical. This study presents an AFC-DRL framework that addresses these challenges by integrating a GPU-accelerated CFD solver, enabling efficient data collection for DRL training and allowing this methodology to be applied to more complex cases. Initially tested on a turbulent boundary layer separation at Reynolds number $Re_\tau=180$ \cite{Font2025} and further validated with a three-dimensional cylinder at $Re = 100$ \cite{Montala2024}, in the present work we propose to advance AFC-DRL applicability to more complex flow scenarios, such as three-dimensional separated flows over wings.

\section{Methodology}

\subsection{DRL set-up}

\begin{figure}[h]
	\begin{center}
	\includegraphics[width=0.9\linewidth]{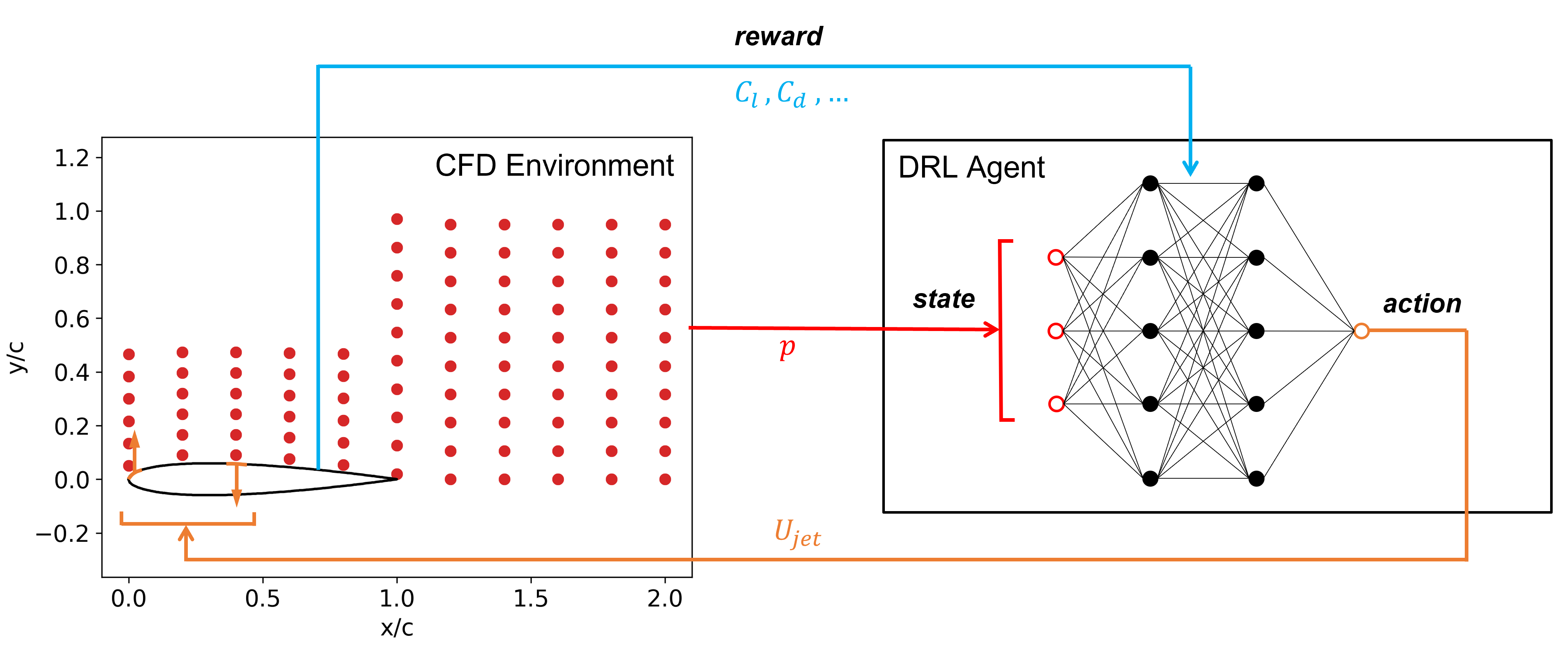}
	\caption{\label{fig:DRL_framework} CFD-DRL set-up.}
	\end{center}
\end{figure}

As illustrated in Fig. \ref{fig:DRL_framework}, the DRL framework consists of two main components: The environment, which is a computational fluid dynamics (CFD) simulation, and the agent, implemented as a neural network (NN). The NN receives as input the state of the flow field, represented by some sensors monitoring a flow variable such as the pressure, and outputs the action to be applied back into the environment, in this case, the jet velocity. The agent undergoes a training process based on trial-and-error learning to optimise the control strategies. Therefore, the CFD solver and the agent not only exchange states and actions but also a reward. The reward function is designed to guide the agent toward optimal actuation strategies. The Proximal Policy Optimization (PPO) algorithm \cite{Schulman2017} is employed to refine the agent’s policy over multiple episodes.

The environment is modelled using the SOD2D solver \cite{Gasparino2024}, a GPU-enabled spectral element method (SEM) code developed at the Barcelona Supercomputing Center (BSC). This solver is designed for high-fidelity simulations and efficiently solves the filtered incompressible Navier-Stokes equations (LES). Its scalability and computational efficiency on distributed accelerator-based architectures make it well-suited for DRL applications, where rapid CFD data generation is crucial for the training.

On the other hand, the DRL agent is implemented using the TF-Agents Python library \cite{Guadarrama2018} and communicates with the CFD solver through a Redis in-memory database, managed via SmartSim \cite{Partee2022}. This setup effectively resolves the "two-language" problem by enabling low-overhead interaction between the Fortran-based CFD solver and the Python-based DRL agent.

\subsection{Case Configuration}
The flow past the three-dimensional NACA0012 wing section at the Reynolds number $Re = 1,000$ and angle of attack AoA $= 20^\circ$ is studied. The Reynolds number is defined using the fluid density $\rho$, the inflow velocity $U_\infty$, the wing chord $c$ and the fluid dynamic viscosity $\mu$ as $Re = \rho U_\infty c/ \mu$. The domain extends $L_x = 32c$, $L_y = 30c$ and $L_z = 3c$ in the streamwise, crosswise and spanwise directions, respectively, with the wing located approximately in the middle of the domain. At the inlet, a constant freestream velocity $U_\infty$ is imposed, while zero-gradient conditions with constant pressure are applied at the outlet. The no-slip condition is enforced on the wing walls, and periodic boundary conditions are applied in the spanwise direction.

When AFC is activated, two sets of three actuators are positioned along the wing's spanwise direction. Each set consists of two jets: one near the leading edge ($x/c = 0.01$) and the other further downstream ($x/c = 0.4$). The two actuators operate with opposite mass flow rates, ensuring instantaneous mass conservation, i.e. $U_{jet, front} = -U_{jet, rear}$. Consequently, the NN only predicts the actuation for the front jet. Each actuator spans a width of $L_{jet} = 1c$, and the agent is restricted to exploring values within the range $U_{jet}/U_\infty = [-1.125, 1.125]$.

\begin{figure}[h]
	\begin{center}
	\includegraphics[width=1.0\linewidth]{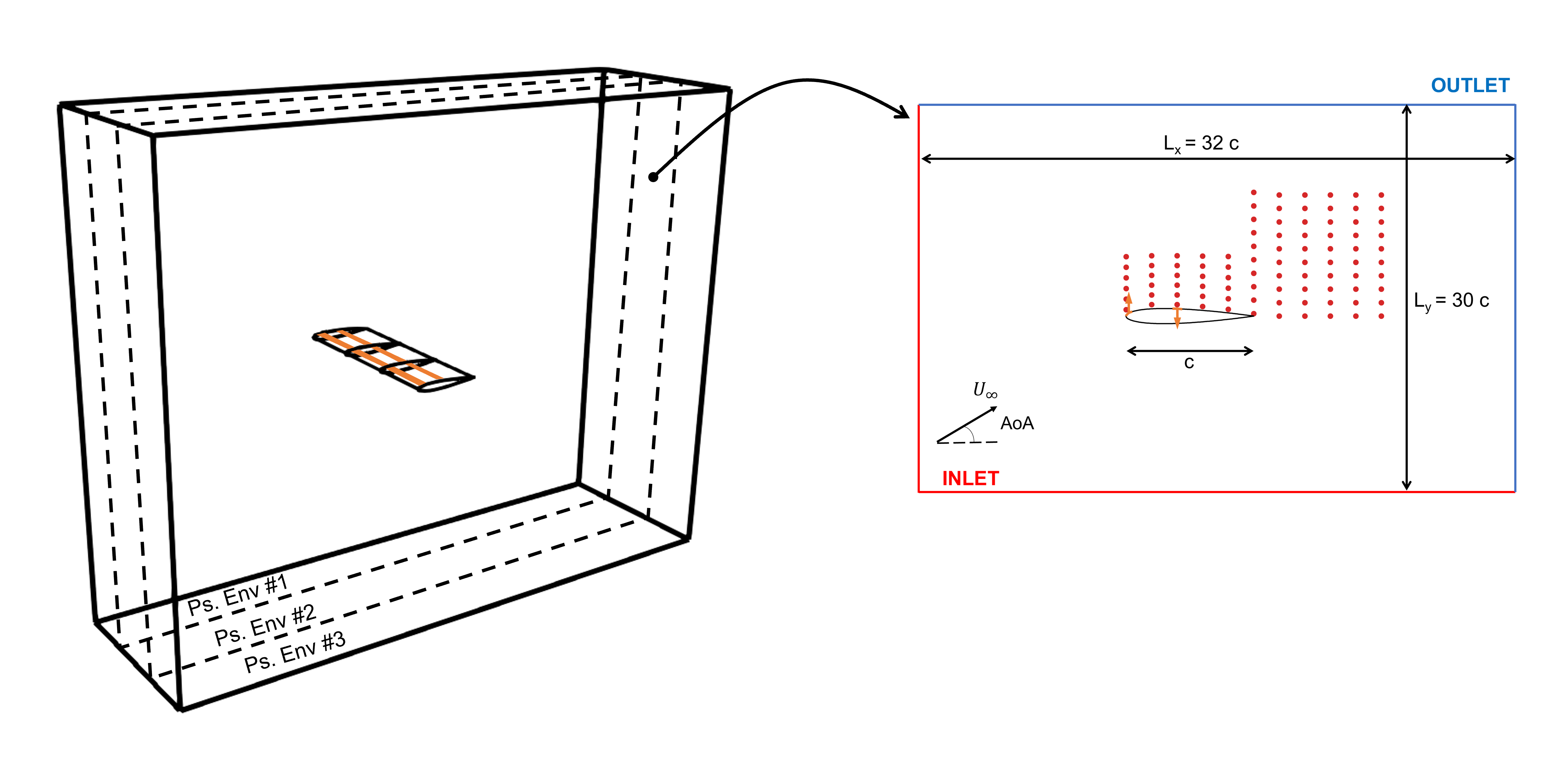}
	\caption{\label{fig:Domain} Configuration of the domain. The left image illustrates the general setup, while a side view of a single pseudo-environment is shown on the right.}
	\end{center}
\end{figure}

Each set of two actuators defines a pseudo-environment, as illustrated in Fig. \ref{fig:Domain}. The entire domain is therefore divided into three pseudo-environments, each with a width of $L_{jet}$. Moreover, ten different CFD simulations are executed in parallel, allowing the DRL agent to collect multiple experiences simultaneously, and applying the multi agent reinforcement learning (MARL) framework \cite{Belus2019}. This was also previously done in Suárez et al. \cite{Suarez2025a, Suarez2025b} and Font et al. \cite{Font2025}, speeding up the training. In total, 30 trajectories (3 pseudo-environments x 10 CFD simulations) are collected after each action (batch size). The duration of an episode $T_{eps}$ corresponds to six vortex-shedding cycles of the baseline scenario. Within each episode, 120 actions are applied, meaning that the duration of each action is $T_{act} = T_{eps}/120$. During this period, an exponential smoothing is employed from the previous to the next predicted action.

For the state, each pseudo-environment includes a slice in the z-middle plane of 90 witness points. The location of these witness points is shown in Fig. \ref{fig:Domain}. However, the input layer of the NN has a size of 90 x 3 = 270, as it incorporates information from the two neighbouring pseudo-environment witness slices as well. The NN architecture consists of two hidden layers, each with 512 neurons.

The reward function for training the model is given in Eq. \ref{eq:reward_1}. The first term rewards the reduction of the drag coefficient $C_d = d/(1/2 \rho U_\infty S)$ with respect to the baseline scenario $C_{d,b}$; while the second penalizes the lift coefficient oscillations $C_l = l/(1/2 \rho U_\infty S)$, with $\alpha$ being a weighting factor to adjust the importance of each term and $C_{l,avg}$ the accumulated mean lift coefficient throughout the episode. Here, the reference surface $S$ is defined as $S = L_{jet} c$, based on the wing chord $c$ and the spanwise length of the jet (or pseudo-environment) $L_{jet}$. The drag $d$ and lift $l$ forces correspond to  the aerodynamic force components in the streamwise and cross-stream directions relative to the freestream velocity 
$U_\infty$.

Finally, the reward applied to the model accounts for both local and global performance, as expressed in Eq. \ref{eq:reward_2}. The weighting factor $\beta$ adjusts the importance of the local reward $r_i$ versus the mean rewards of all pseudo-environments $\sum^{n_{jets}}_{j=1} r_j$. In the present study, the weighting factors in Eq. \ref{eq:reward_1} and Eq. \ref{eq:reward_2} are set to $\alpha = 0.3$ and $\beta = 0.8$, respectively.

\vskip-.6cm
\begin{eqnarray}
\label{eq:reward_1} r_i = (C_{d,b} - C_d) - \alpha |C_l - C_{l, avg}| \\
\label{eq:reward_2} R_i = \beta r_i + (1-\beta)/n_{jets}\sum_{j=1}^{n_{jets}} r_j
\end{eqnarray}

\section{Results}

\begin{figure}[h]
    \centering
    \begin{subfigure}{0.4\textwidth}
    \centering
    \includegraphics[width=\linewidth]{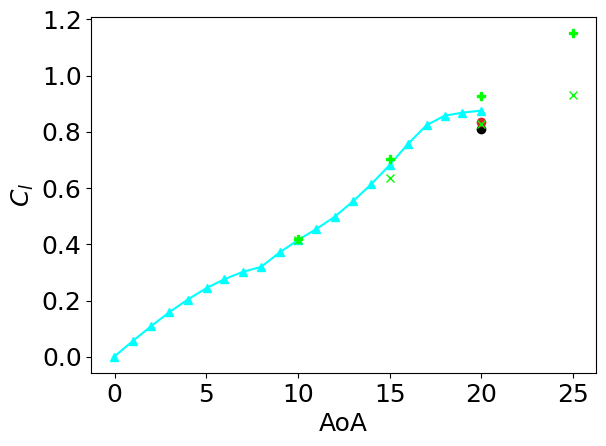}
    \caption{}
    \label{fig:Cl_alpha}
    \end{subfigure}
    \begin{subfigure}{0.4\textwidth}
    \centering
    \includegraphics[width=\linewidth]{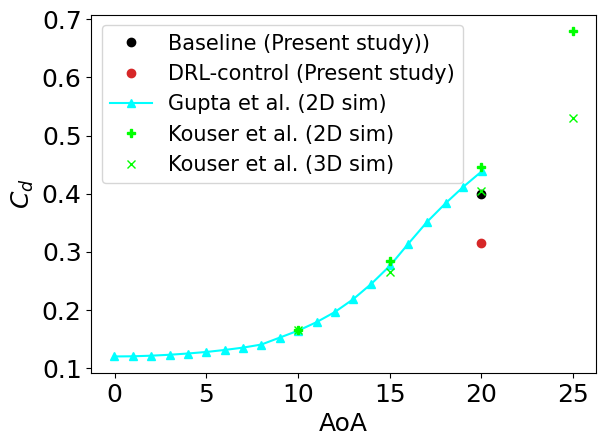}
    \caption{}
    \label{fig:Cd_alpha}
    \end{subfigure} \\
    \begin{subfigure}{0.4\textwidth}
    \centering
    \includegraphics[width=\linewidth]{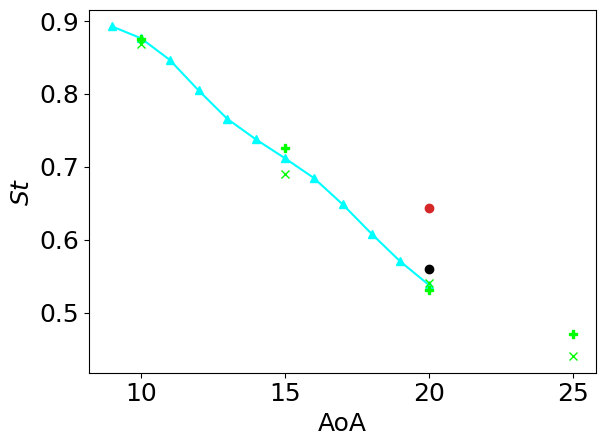}
    \caption{}
    \label{fig:St_alpha}
    \end{subfigure}
    \begin{subfigure}{0.4\textwidth}
    \centering
    \includegraphics[width=\linewidth]{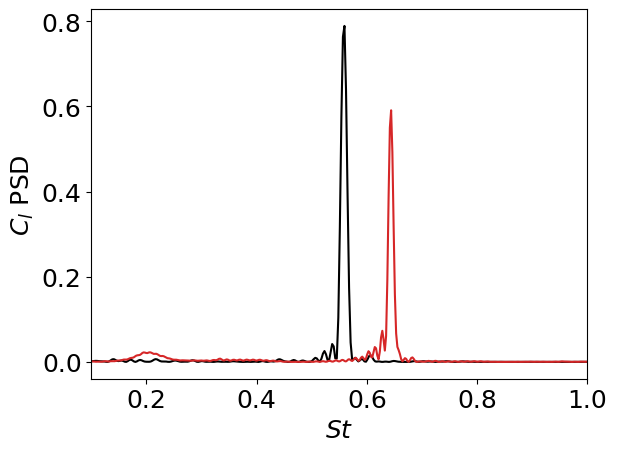}
    \caption{}
    \label{fig:FT_Cl}
    \end{subfigure}
    \caption{(a) Lift coefficient, (b) drag coefficient and (c) Strouhal number of the vortex shedding  as a function of the angle of attack (AoA) [$^\circ$], compared with the literature \cite{Gupta2023, Kouser2021}; (d) Energy spectrum of the lift coefficient.}
    \label{fig:coeff_alpha}
\end{figure}

As demonstrated by Gupta et al. \cite{Gupta2023}, for the Reynolds number studied in this work, i.e., $Re=1,000$, vortex shedding begins to appear for angles of attack greater than AoA$ > 8^\circ$, while three-dimensional flow structures emerge at approximately AoA $ > 13^\circ$. This explains the discrepancies observed in the literature between two-dimensional and three-dimensional cases in terms of lift and drag coefficients (see Fig. \ref{fig:Cl_alpha} and \ref{fig:Cd_alpha}, respectively), as well as variations in the Strouhal number of vortex shedding (see Fig. \ref{fig:St_alpha}).

\begin{figure}[h]
    \centering
    \begin{subfigure}{0.4\textwidth}
    \centering
    \includegraphics[width=\linewidth]{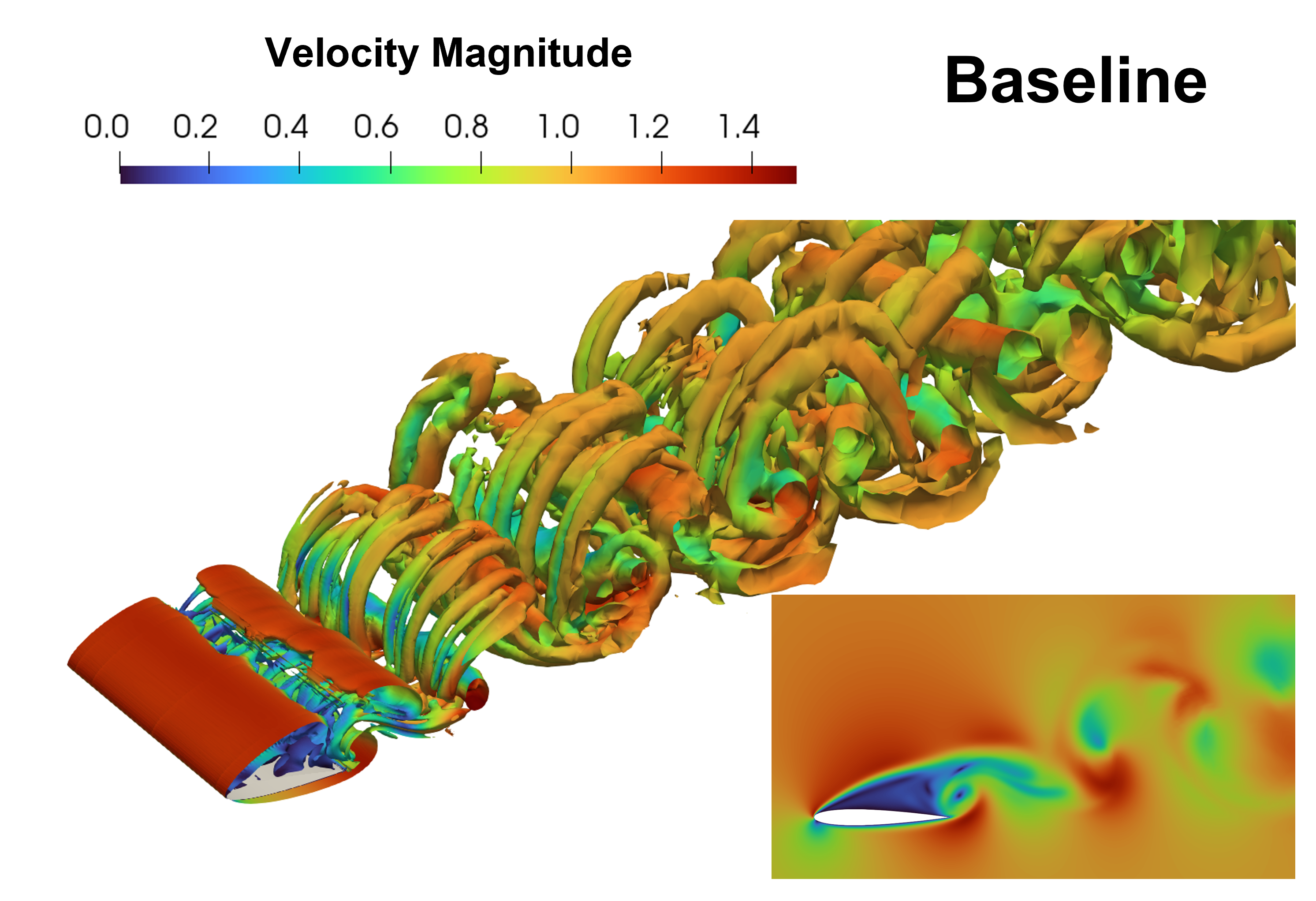}
    \caption{}
    \label{fig:Qs_base}
    \end{subfigure}
    \begin{subfigure}{0.4\textwidth}
    \centering
    \includegraphics[width=\linewidth]{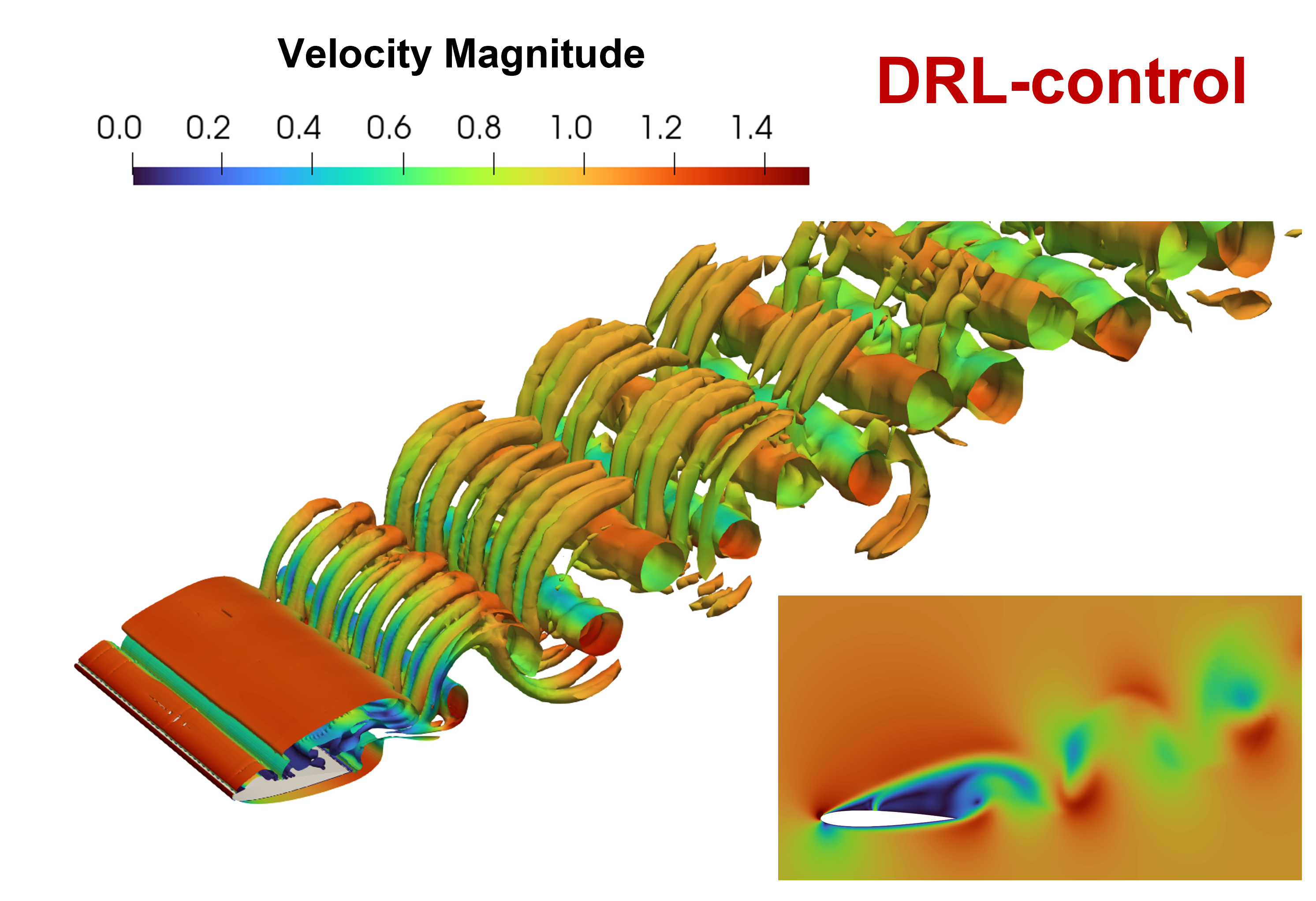}
    \caption{}
    \label{fig:Qs_DRL}
    \end{subfigure}\\
    \begin{subfigure}{0.4\textwidth}
    \centering
    \includegraphics[width=\linewidth]{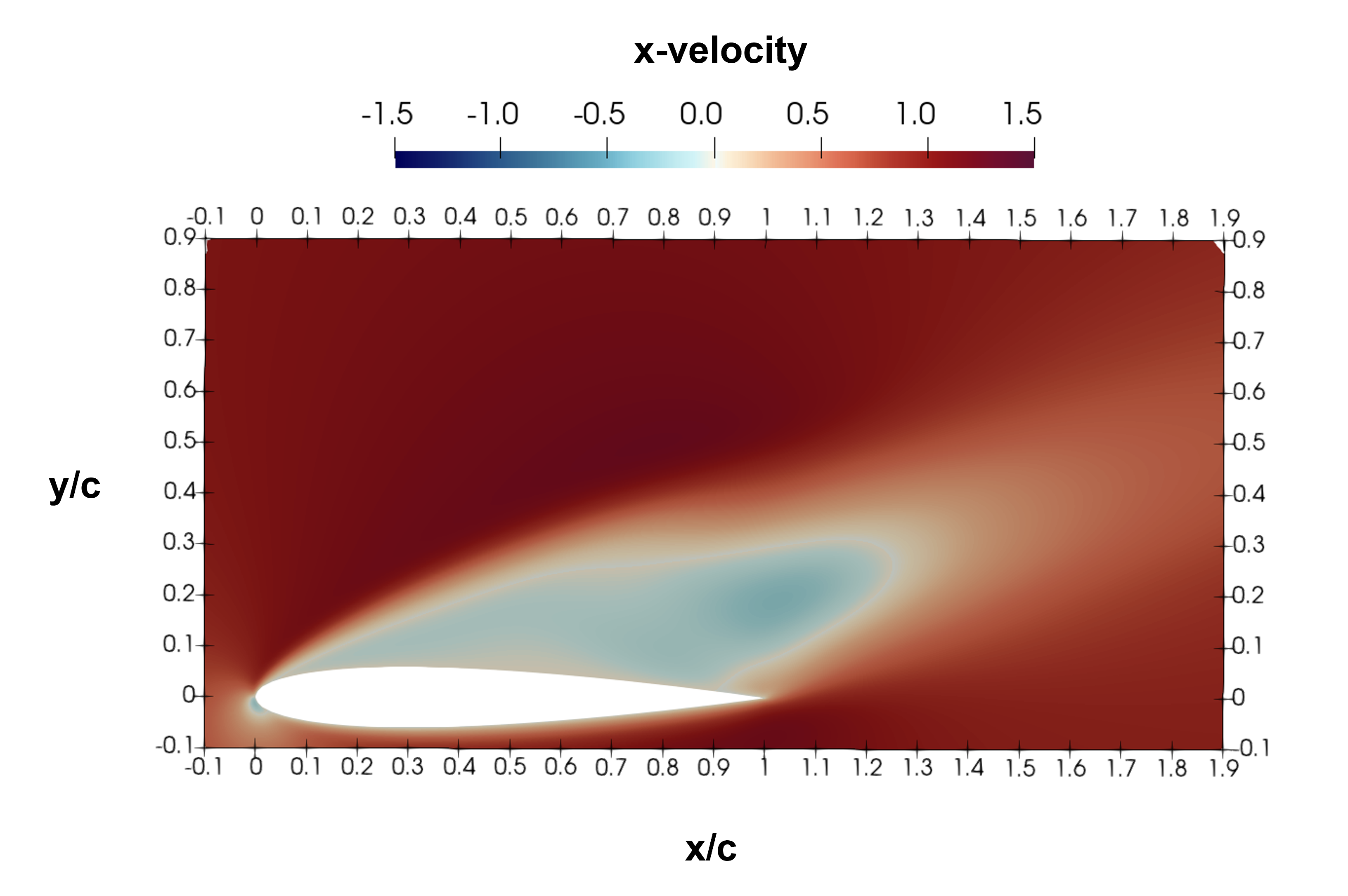}
    \caption{}
    \label{fig:xVel_base}
    \end{subfigure}
    \begin{subfigure}{0.4\textwidth}
    \centering
    \includegraphics[width=\linewidth]{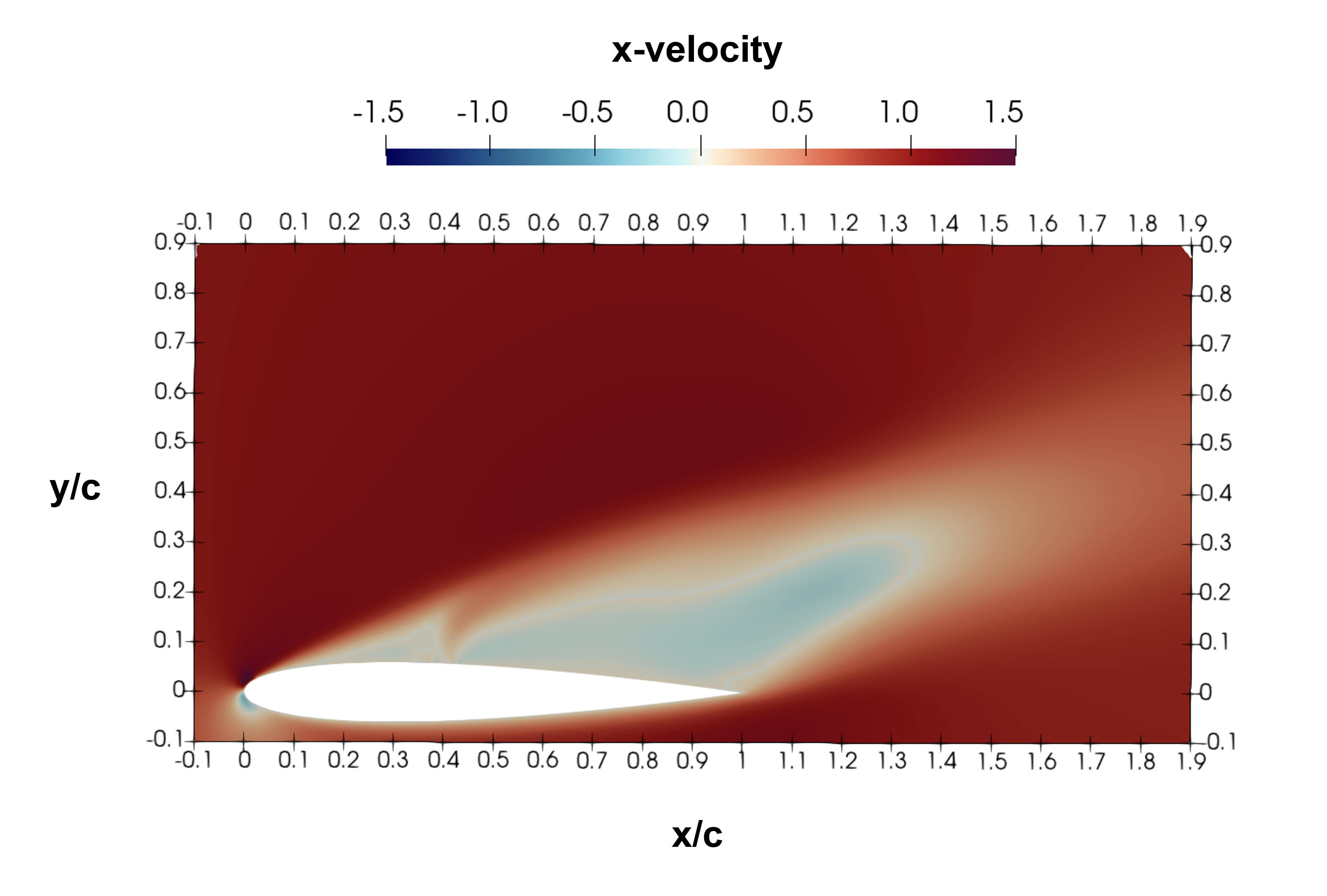}
    \caption{}
    \label{fig:xVel_DRL}
    \end{subfigure}
    \caption{Q-criterion iso-contours colored by velocity magnitude, along with the instantaneous velocity field in the middle z-plane (top); and time-averaged x-velocity contours (bottom) for the baseline (a, c) and DRL-control (b, d) scenarios.}
    \label{fig:contours}
\end{figure}

In the present study, the angle of attack is fixed at AoA = $20^\circ$, where both vortex shedding and three-dimensional flow structures are present. The baseline case demonstrates good agreement with existing three-dimensional results in the literature, as shown in Fig. \ref{fig:coeff_alpha}. Instantaneous contours of the baseline scenario are depicted in Fig. \ref{fig:Qs_base}, clearly illustrating the von Kármán vortex street and the three-dimensional features of the wake. As the angle of attack increases, the large separation on the wing's suction side expands, which triggers the development of three-dimensional flow features. Consequently, significant flow separation is observed in the studied case, as shown in the averaged x-velocity contours in Fig. \ref{fig:xVel_base}, where a distinct region of negative velocity is visible. The main objective of this study is to control this separation by training a DRL model that learns a complex AFC actuation that reduces the baseline drag coefficient, together with the amplitude of the lift oscillations, as expressed in the reward function described in Eq. \ref{eq:reward_1}.

\begin{figure}[h]
    \centering
    \begin{subfigure}{0.32\textwidth}
    \centering
    \includegraphics[width=\linewidth]{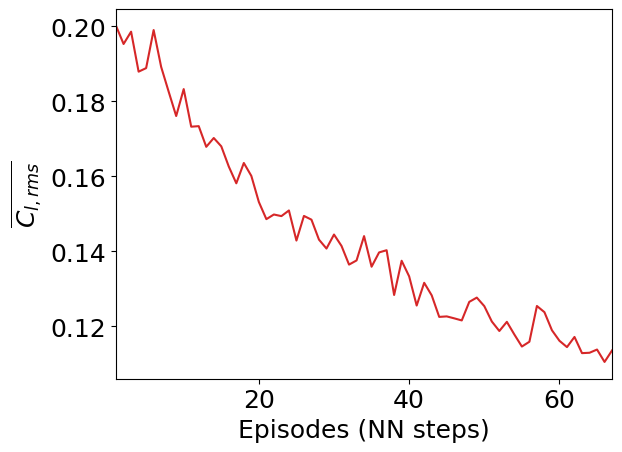}
    \caption{}
    \label{fig:Clrms_train}
    \end{subfigure}
    \begin{subfigure}{0.32\textwidth}
    \centering
    \includegraphics[width=\linewidth]{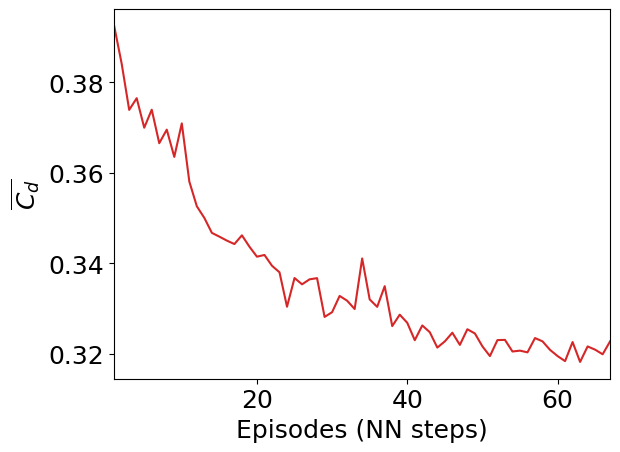}
    \caption{}
    \label{fig:Cd_train}
    \end{subfigure}
    \begin{subfigure}{0.32\textwidth}
    \centering
    \includegraphics[width=\linewidth]{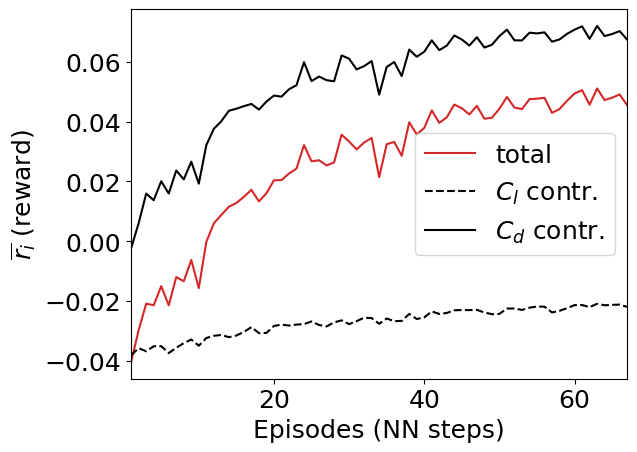}
    \caption{}
    \label{fig:reward_train}
    \end{subfigure}
    \caption{Evolution of the (a) root-mean-square lift coefficient, (b) drag coefficient, and (c) reward during the DRL training, averaged over the last 5 flow-throughs of each episode across the 30 environments.}
    \label{fig:training}
\end{figure}

Fig. \ref{fig:reward_train} shows how the mean reward ($\overline{r_i}$) evolves over the training episodes, i.e., NN steps, where the contributions from the drag and lift terms of Eq. \ref{eq:reward_1} are plotted separately as well. The mean reward is calculated by averaging the rewards ($r_i$) across the 30 pseudo-environments that are run in parallel over the last 5 time units of each episode. From the figure, it is clear that the agent learns a strategy that effectively reduces both the lift coefficient oscillations (represented by the root-mean-square) and the drag coefficient, as can be observed in Fig. \ref{fig:Clrms_train} and Fig. \ref{fig:Cd_train}, respectively. These figures show how both values decrease progressively as the training advances, leading to a higher total reward. By the end of the training, the reward reaches a plateau, and the training is stopped after 67 episodes. To assess the learned actuation strategy, the model is then run in deterministic mode, applying the learned policy without further exploration.

\begin{figure}[h]
    \centering
    \begin{subfigure}{0.4\textwidth}
    \centering
    \includegraphics[width=\linewidth]{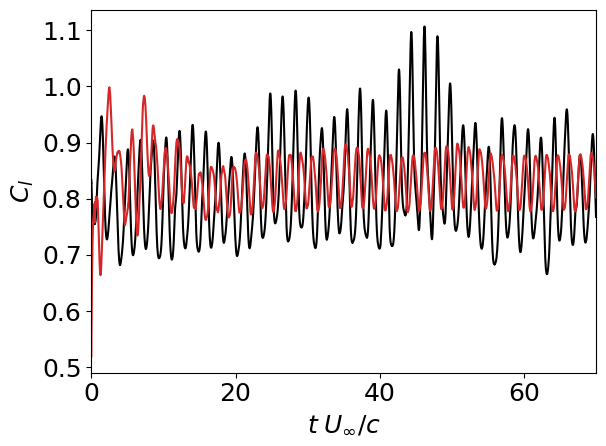}
    \caption{}
    \label{fig:Cl_time}
    \end{subfigure}
    \begin{subfigure}{0.4\textwidth}
    \centering
    \includegraphics[width=\linewidth]{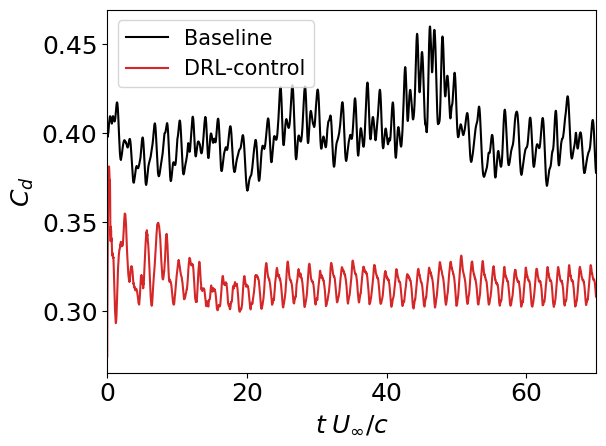}
    \caption{}
    \label{fig:Cd_time}
    \end{subfigure}
    \caption{(a) Lift coefficient and (b) drag coefficient temporal signals.}
    \label{fig:coeff_time}
\end{figure}

\begin{figure}[h]
    \centering
    \begin{subfigure}{0.4\textwidth}
    \centering
    \includegraphics[width=\linewidth]{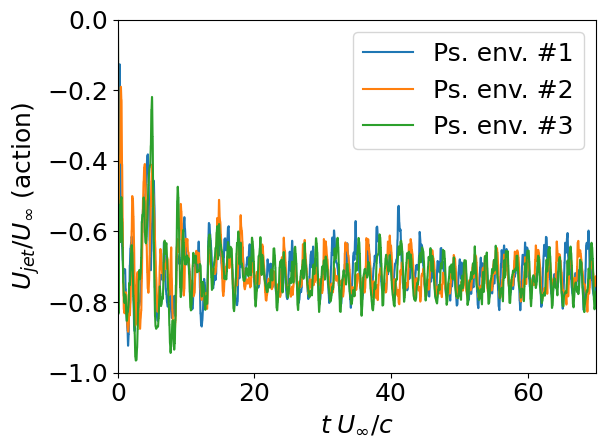}
    \caption{}
    \label{fig:jetVel_time}
    \end{subfigure}
    \begin{subfigure}{0.4\textwidth}
    \centering
    \includegraphics[width=\linewidth]{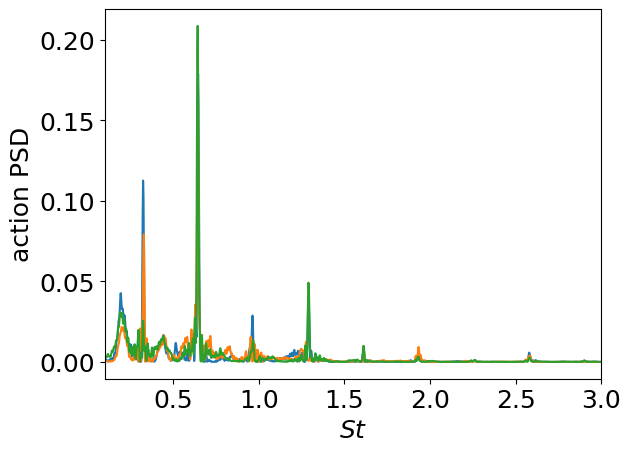}
    \caption{}
    \label{fig:FT_jetVel}
    \end{subfigure}
    \caption{(a) Instantaneous jet velocities applied in each pseudo-environment, and (b) energy spectrum of the jet velocities.}
    \label{fig:actions}
\end{figure}

Fig. \ref{fig:coeff_time} compares the instantaneous lift and drag coefficients obtained with the learned DRL-based control against the baseline temporal signals. These results illustrate the effectiveness of the control strategy in significantly reducing lift oscillations and drag. At the beginning of the actuation, a transient period occurs as the DRL agent gradually adjusts and controls the flow, eventually reaching a new statistical state after about 20 flow-throughs. As shown in the plots, there are small variations in the mean lift coefficient compared to the baseline case, but the root-mean-square of the signal has been noticeably reduced ($\Delta C_{l,rms} = -124 \%$). Additionally, a substantial mean drag reduction is observed ($\Delta C_d = -21 \%$), which aligns with the objectives explicitly defined in the reward function.

The mean drag and lift coefficients are further compared against the uncontrolled case in Fig. \ref{fig:coeff_alpha}. This figure also presents the Strouhal number of the vortex shedding, showing that the flow dynamics have been slightly accelerated by the actuation, with the shedding frequency increasing from St$=0.56$ to $0.64$.  The energy spectrum of the lift coefficient is depicted in Fig. \ref{fig:FT_Cl}, where a distinct peak can be seen at the vortex-shedding frequencies of each case.

The instantaneous actions applied by each pseudo-environment are shown in Fig. \ref{fig:jetVel_time}, along with their corresponding energy spectrum in Fig. \ref{fig:FT_jetVel}. These actions correspond to the jets located at the leading edge, while the rear jets are constrained to apply signals of the opposite sign. The energy spectrum shows that the dominant frequency of the actions precisely matches the vortex-shedding frequency, with the signals of all the pseudo-environments exhibiting a prominent peak at St$=0.64$. Furthermore, an analysis of the instantaneous jet velocities indicates that the DRL agent has learned a strategy where the front jet maintains a negative mean value (suction), making the rear jet continuously blow (positive mean jet velocity). Additionally, it is interesting to note that the instantaneous jet velocities vary across pseudo-environments to collectively establish a regular pattern of vortical structures along the spanwise direction. 

This is evident in the vortical structures visualized in Fig. \ref{fig:Qs_DRL} by means of the Q-criterion iso-contours, which are compared against the baseline structures in Fig. \ref{fig:Qs_base}. These images show that the shear layer originating from the wing’s leading edge extends further in the streamwise direction, delaying the onset of vortex shedding and resulting in a more organized wake. This, in turn, reduces the three-dimensionality of the flow, leading to a more regular periodicity of the flow with fewer temporal and spatial scales.

This behaviour is also reflected in the lift and drag coefficients of the baseline case, as shown in Fig. \ref{fig:coeff_time}. In the uncontrolled scenario, some chaotic fluctuations are observed due to the high angle of attack and the presence of strong pressure gradients, leading to sudden increases in the coefficients. However, in the controlled scenario, these spurious variations are entirely suppressed, resulting in nearly constant amplitude oscillations in both lift and drag coefficients.

Finally, the averaged flow separation can be observed in the negative x-velocity regions shown in Fig. \ref{fig:xVel_base} and Fig. \ref{fig:xVel_DRL} for the uncontrolled and controlled cases, respectively. These figures clearly demonstrate how the actuation effectively reduces the flow separation, further explaining the improved lift and drag coefficients obtained through AFC.

\section{Conclusions}

This study applies deep reinforcement learning (DRL) for active flow control (AFC) to a three-dimensional wing at a high angle of attack (AoA = $ 20^\circ$), where vortex shedding and three-dimensional flow structures are fully developed. The DRL framework is integrated with the GPU-accelerated CFD solver SOD2D, allowing for efficient training and rapid experience collection across multiple pseudo-environments.

The results demonstrate that the DRL-based control strategy effectively reduces both drag and lift oscillations by dynamically adjusting the actuation. Specifically, the root-mean-square of the lift coefficient is reduced by 124\%, while the mean drag coefficient decreases by 21\%, aligning well with the reward function objectives. These improvements are achieved by controlling the shear layer development and delaying vortex shedding, leading to a more structured wake with reduced three-dimensional flow features.

Moreover, the learned actuation strategy exhibits a strong correlation with the vortex-shedding frequency (St $= 0.64$), indicating that the DRL agent has successfully identified an optimal periodic forcing mechanism. This mechanism includes additional frequency components that enable further adjustments to the real-time flow conditions. The results also highlight the role of multiple pseudo-environments working together to establish a coherent actuation pattern along the spanwise direction, leading to a regular and periodic wake structure.

These findings reinforce the potential of DRL in discovering advanced AFC strategies beyond traditional periodic forcing techniques. Furthermore, the scalability of the current framework suggests that it can be extended to more complex aerodynamic configurations at higher Reynolds numbers, paving the way for data-driven control strategies in real-world aerodynamic applications.

\section*{Acknowledgments}

This research has received financial support from the Ministerio de Ciencia e Innovacion of Spain (PID2023-150408OB-C21 and PID2023-150408OB-C22). Simulations were conducted with the assistance of the Red Española de Supercomputación (RES) and the EuroHPC JU, who granted us computational resources at the HPC facilities of MareNostrum V at Barcelona Supercomputing Cente (IM-2024-2-0004 and EHPC-REG-2024R01-038, respectively). Authors also extend their gratitute to the Agència de Gestió d'Ajuts Universitaris i de Recerca (AGAUR) for supporting the research group Large-scale Computational Fluid Dynamics (2021 SGR 00902) and the Turbulence and Aerodynamics Research Group (2021 SGR 01051). Ricard Montalà express also gratitude to AGAUR for awarding the FI-SDUR grant (2022 FISDU 00066), which supports his doctoral studies. Finally, Ricardo Vinuesa acknowledges financial support from ERC grant no.2021-CoG-101043998, DEEPCONTROL. Views and opinions expressed are however those of the author(s) only and do not necessarily reflect those of the European Union or the European Research Council. Neither the European Union nor the granting authority can be held responsible for them.

\end{document}